\documentclass[graybox]{svmult}

\usepackage{mathptmx}       
\usepackage{helvet}         
\usepackage{courier}        
\usepackage{type1cm}        
\usepackage{makeidx}         
\usepackage{graphicx}        
\usepackage{multicol}        
\usepackage[bottom]{footmisc}

\usepackage{multirow}
\usepackage{longtable}
\usepackage{bigstrut}

\makeindex             

\begin{document}

\title*{Geoneutrinos and reactor antineutrinos at SNO+}
\author{M Baldoncini$^{1,2}$, V Strati$^{1,3}$, S A Wipperfurth$^{4}$, G Fiorentini$^{1,2,3}$, F 
Mantovani$^{1,2}$, W F McDonough$^{4}$ and B Ricci$^{1,2}$}

\authorrunning{Baldoncini et al.}

\institute{
$^{1}$ University of Ferrara, Department of Physics and Earth Sciences, Ferrara, Italy,
\and 
$^{2}$ INFN, Ferrara Section, Ferrara, Italy
\and 
$^{3}$ INFN, Legnaro National Laboratories, Legnaro, Italy
\and
$^{4}$ Department of Geology, University of Maryland, College Park, Maryland, USA}

%
%
\maketitle

\abstract{
In the heart of the Creighton Mine near Sudbury (Canada), the SNO+ detector is foreseen to observe almost in equal 
proportion electron antineutrinos produced by U and Th in the Earth and by nuclear reactors. SNO+ will be the first 
long baseline experiment to measure a reactor signal dominated by CANDU cores ($\sim$55\% of the total reactor signal), 
which generally burn natural uranium. Approximately 18\% of the total geoneutrino signal is generated by the U and Th 
present in the rocks of the Huronian Supergroup-Sudbury Basin: the 60\% uncertainty on the signal produced by this 
lithologic unit plays a crucial role on the discrimination power on the mantle signal as well as on the geoneutrino 
spectral shape reconstruction, which can in principle provide a direct measurement of the Th/U ratio in the Earth.
}

\section{Introduction}
\label{sec:1}

Designed as a retrofit of the former Sudbury Neutrino Observatory (SNO) at SNOLAB, SNO+ is a multipurpose kiloton-scale 
liquid scintillation detector aimed at performing low energy neutrino physics measurements. Thanks to an overburden of 6 
km water equivalent and to a very low background, the SNO+ detector can reach several physics goals, including the 
observation of electron antineutrinos produced by the Earth and by nuclear reactors via the Inverse Beta Decay (IBD) 
reaction \cite{chen_2006}.

Geoneutrinos produced in beta minus decays along the $^{238}$U and $^{232}$Th decay chains provide an exceptional 
insight into the Earth's interior, allowing for the determination of the heat-producing element abundances and hence of 
the total radiogenic heat power of the planet. The SNO+ detector is located in the Superior Province (Ontario, Canada), 
one of the Earth's largest Archean's cratons, characterized by a thick ($\sim$ 42 km) continental crust, which gives 
rise to a sizable geoneutrino crustal signal rate \cite{huang_2014}. In this framework, nuclear reactors are the 
most severe source of background as there is a significant overlap between the reactor antineutrino energy spectrum and 
the geoneutrino one at low energies. The Bruce, Pickering and Darlington power stations, which are respectively 240 km, 
340 km and 350 km far from the SNO+ site, host globally 18 operating cores for a total thermal power of approximately 43 
GW \cite{baldoncini_2015}.

The primary aim of this study is predicting the expected geoneutrino and reactor signals at SNO+ on the basis of 
existing reference Earth and reactor models. Since these predictions are affected by some degree of uncertainties, the 
purpose is to highlight priorities for future refinements.

\begin{figure}
 \center
 \includegraphics[width=0.8\textwidth]{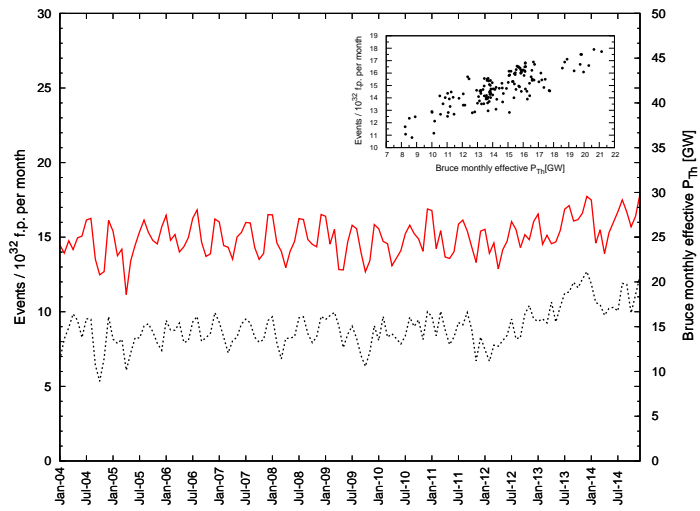}
 \caption{SNO+ reactor signal and Bruce power station's effective thermal power. The SNO+ signal in 
the FER (red solid line, left y axis) and the Bruce power station's effective thermal power (black dashed line, right y 
axis) are reported on a monthly time scale from January 2004 to December 2014.  In the top-right box a scatter plot of 
the total reactor signal in the FER and the Bruce power station's effective thermal power is shown.}
\label{bruce_fig}
\end{figure}

\section{Geoneutrinos}

The prediction of the geoneutrino signal at SNO+ is based on the modeling of the distribution and amount of U and Th in 
the Earth's reservoirs. The continental crust, despite accounting for approximately 0.5\% of the Earth's mass, is the 
main reservoir of U and Th and generates 75\% of the total geoneutrino signal expected at SNO+ \cite{huang_2013}. 
For this reason a deep understanding of the continental crust, in particular the region immediately surrounding the 
detector, is mandatory to evaluate the geoneutrino signal and its uncertainties. These studies can be performed both via 
3D geochemical and geophysical crustal models \cite{huang_2014} and via heat balance models based on the combination 
of the Moho heat flux and of the crustal heat production \cite{phaneuf_2014}.

The local crust of SNO+, i.e. six 
2$^{\odot}$  $\times$ 2$^{\odot}$ 
tiles centered at the detector location, is modeled in \cite{huang_2014} 
and is based on integrating regional geological, geophysical, and geochemical data. The geoneutrino signal from the 
local crust is predicted to be 15.6$^{+5.3}_{-3.4}$ TNU. A detailed analysis of the geoneutrino signal contribution 
from the different lithologic units of the local crust reveals that the Huronian Supergroup-Sudbury Basin (HS-SB), with 
a signal of 7.3$^{+5.0}_{-3.0}$ TNU, is the major source of signal as well as of uncertainty concerning the local 
contribution to the expected geoneutrino signal. The $\pm$1$\sigma$ range (4.3 - 12.3 TNU) of the geoneutrino signal 
produced by the HS-SB is comparable with the full range of the expected mantle signal (2 - 19 TNU) 
\cite{fiorentini_2012, 
sramek_2013}. With the perspective of inferring the mantle geoneutrino signal by subtracting the estimated crustal 
contribution to the total  \cite{fiorentini_2012}, improving the HS-SB modeling is mandatory.

\begin{figure}
\centering
 \includegraphics[width=0.8\textwidth]{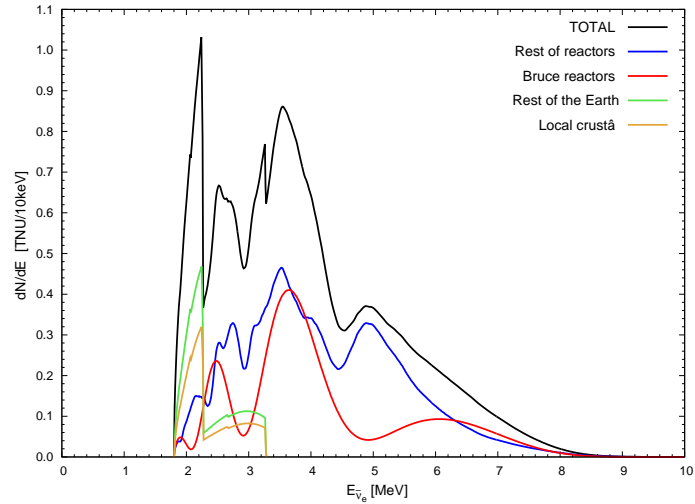}
 \caption{Geoneutrino and reactor antineutrino spectra at SNO+. The geoneutrino spectrum is 
subdivided into the local crust (brown line) and rest of the Earth (green line) components. The reactor spectrum is 
split into the contributions generated by the Bruce power station (red line) and by the rest of reactors (blue line). 
The overall antineutrino spectrum at SNO+ (black line) is also shown.}
\label{spectra_fig}
\end{figure}

In Fig.{\ref{spectra_fig} we report the geoneutrino spectrum subdivided into two components, generated by the U 
and Th in the local crust and in the rest of the Earth. We took into account the antineutrino survival probability 
adopting the oscillation parameters reported in \cite{huang_2014}. In Table \ref{signal_table}} we summarize the 
four main geoneutrino signal components at SNO+, corresponding to the local crust, rest of the crust, continental 
litospheric mantle and mantle contributions.

\begin{table}
\centering
\caption{Geoneutrino and reactor antineutrino signals at SNO+ reported in TNU. The total geoneutrino 
signal is the sum of the contributions from the local crust, the rest of the crust, the continental litospheric mantle 
and the mantle. The reactor signal is given separately for the LER and FER. The total reactor signal is the sum of the 
contributions generated by the Bruce power station and by the rest of the reactors.}

\vspace{10pt}
\renewcommand{\arraystretch}{1.1}
\begin{tabular}{c c c c c}
\hline \hline
\multicolumn{2}{c}{Geoneutrinos}& \multicolumn{3}{c}{Reactor antineutrinos}\\
\cline{1-5}

\multicolumn{1}{c}{}&\multicolumn{1}{c}{LER [TNU]}&&LER [TNU]&FER [TNU]  \bigstrut[t] \\
\cline{1-5}
\multicolumn{1}{c}{Local crust}&\multicolumn{1}{c}{15.6$^{+5.3}_{-3.4}$}&Bruce 
reactors&17.3$^{+1.0}_{-0.7}$&73.7$^{+2.0}_{-1.8}$ \bigstrut[t]\\
\cline{1-5}
\multicolumn{1}{c}{Rest of the crust}&\multicolumn{1}{c}{15.1$^{+2.8}_{-2.4}$}&\multirow{3}{*}{Rest 
of reactors}&\multirow{3}{*}{
31.2$^{+0.9}_{-0.8}$}&\multirow{3}{*}{118.9$^{ +2.8}_{-2.6}$} \bigstrut[t] \\
\multicolumn{1}{c}{Continental litoshperic mantle} & \multicolumn{1}{c}{2.1$^{+2.9}_{-1.2}$}& & & \bigstrut[t] \\
\multicolumn{1}{c}{Mantle} &\multicolumn{1}{c}{9} & && \bigstrut[t] \\
\cline{1-5}
\multicolumn{1}{c}{TOTAL}& \multicolumn{1}{c}{40$^{+6}_{-4}$}&TOTAL&48.5$^{+1.8}_{-1.5}$&192.6$^{+4.7}_{-4.4}$ 
\bigstrut[t] \\
\hline \hline

\end{tabular}
\label{signal_table}
\end{table}

\section{Conclusions and perspectives}
As the antineutrino signal in the LER at SNO+ is expected to be generated by nuclear reactors and by the Earth in a 
ratio of $\sim$1.2, a detailed characterization of both the geoneutrino and the reactor antineutrino sources is 
compulsory in terms of both absolute values and uncertainties.

A multitemporal analysis of the expected reactor signal at SNO+ over a time lapse of 10 years (January 2004 – December 
2014) reveals that the monthly signal fluctuations associated to different reactors operational condition are on the 
order of 10\% at 1$\sigma$ level. The 18 operating CANDU reactors belonging to the Bruce, Pickering and Darlington 
power stations generate approximately 55\% of the total reactor signal in the LER at SNO+. For this reason, an accurate 
profile of the CANDU fissile isotope inventory over the entire duty cycle is mandatory for the SNO+ experiment. 

The HS-SB is the strongest geoneutrino source among the local crust reservoirs and it is predicted to produce 
7.3$^{+5.0}_{-3.0}$ TNU with respect to a total geoneutrino signal of 40$^{+6}_{-4}$ TNU. 
The compositional heterogeneity of this lithologic unit, which is due to the presence of a mixture of Paleoproterozoic 
sedimentary, metasedimentary and igneous rocks of the Canadian Shield, affects the geoneutrino signal uncertainty on the 
order of 60\%. Therefore, a systematic sampling of the main lithologies of the HS-SB is highly recommended for 
improving the knowledge of the U and Th content of the unit and to put more stringent constraints on the local 
contribution to the geoneutrino signal at SNO+.

\end{document}